\documentclass[a4paper,11pt]{article}
\pdfoutput=1 % if your are submitting a pdflatex (i.e. if you have
\usepackage{jinstpub} % for details on the use of the package, please
\usepackage{lineno}
\usepackage{subcaption}
\usepackage{siunitx}
\DeclareSIUnit\clight{\text{\ensuremath{c}}}
\usepackage{verbatim}
\title{\boldmath Charge collection parameterization of MALTA2, a depleted monolithic active pixel sensor}

\author[a,b,1]{L. Fasselt\note{Corresponding author.}}
\author[c]{P. Behera}%\email{}
\author[a,b]{D. V. Berlea}%\email{}
\author[d]{D. Bortoletto}%\email{}
\author[e]{C. Buttar}%\email{} Craig
\author[c]{T. Chembakan}%\email{}
\author[f]{V. Dao}%\email{} Valerio
\author[c]{G. Dash}%\email{}
\author[g]{S. Haberl}%\email{}
\author[g]{T. Inada}%\email{}
\author[h]{F.K. Isik}%\email{}
\author[c]{P. Jana}%\email{}
\author[i]{X. Li}%\email{}
\author[j]{L. Li}%\email{}
\author[g]{H. Pernegger}%\email{}
\author[g]{P. Riedler}%\email{}
\author[g]{W. Snoeys}%\email{}
\author[g]{C.A. Solans Sánchez}%\email{}
\author[g]{A. Swoboda}%\email{}
\author[h]{I. Turk Cakir}%\email{}
\author[g]{M. van Rijnbach}%\email{}
\author[g]{M. Vázquez Núñez}%\email{}
\author[c]{A. Vijay}%\email{}
\author[g]{J. Weick}%\email{}
\author[a,b]{S. Worm}%\email{}

\affiliation[a]{Deutsches Elektronen-Synchrotron DESY, Zeuthen, Germany}
\affiliation[b]{Humboldt University of Berlin, Berlin, Germany}
\affiliation[c]{Indian Institute of Technology Madras, Chennai, India}
\affiliation[d]{University of Oxford, Oxford, UK}
\affiliation[e]{University of Glasgow, Glasgow, UK}
\affiliation[f]{Stony Brook University, New Yorck, US}
\affiliation[g]{CERN, Geneva, Switzerland}
\affiliation[h]{Ankara University, Ankara, Turkey} %
\affiliation[i]{Los Alamos National Laboratory, Los Alamos, US} %
\affiliation[j]{University of Birmingham, Birmingham, UK} %\orgdiv{School of Physics and Astronomy},

\emailAdd{lucian.fasselt@desy.de}

\abstract{A fast simulation method is presented for a depleted monolithic active pixel sensor, which uses a data driven parameterization of the charge collection and propagation.
This approach provides an efficient alternative to TCAD simulations, particularly for sensors whose proprietary process details - such as doping profiles or implant geometries - are unavailable.
Data was obtained with a MALTA2 sensor fabricated in a \SI{180}{\nano\meter} CMOS imaging technology on \SI{30}{\micro\meter} epitaxial silicon using the MALTA beam telescope at CERN SPS. % and the Edge Transient Current Technique (Edge-TCT).
The model reproduces the measured in-pixel efficiency with high accuracy and enables a realistic yet computationally lightweight analog pixel simulation.
This method will be further employed in optimizing the digital sensor design for applications in high-rate particle tracking and high-granularity calorimetry.}

\keywords{Detector modelling and simulations II (electric fields, charge transport, multiplication and induction, pulse formation, electron emission, etc); Particle tracking detectors; Radiation-hard detectors}

\proceeding{Topical Workshop on Electronics for Particle Physics\\
Rethymno, Crete, Greece\\
6-10 October 2025\\}

\notoc % ToC not mandatory

\begin{document}
\maketitle
\flushbottom

\section{Introduction}
\label{sec:intro}
\setcounter{page}{1}
The simulation of charge collection is essential in sensor design, characterization and detector optimization. 
Depleted monolithic active pixel sensors (DMAPS) with small collection electrodes employ non-uniform electric field configurations, requiring accurate modeling of charge transport and signal formation.
Technology Computer-Aided Design (TCAD) device simulations are widely used in the design stage to predict key characteristics such as capacitance, gain, timing, and detection efficiency \cite{MunkerTCAD}.
They rely on detailed knowledge of doping profiles, silicon resistivity, and implant geometries, allowing the designer to optimize the layout before prototype fabrication.
%While computationally demanding, TCAD remains less costly and time-consuming than manufacturing and testing physical prototypes.
However, a common problem for silicon sensors is that proprietary information is not openly available from the foundry for commercial manufacturing processes.
In this case, TCAD modeling must rely on assumptions and careful systematic parameter variations \cite{SimGenericTCAD}.
Although such studies can qualitatively predict the influence of individual parameters, their accuracy ultimately depends on experimental validation.
In this work, we present an alternative fast simulation approach based on a measurement-driven parameterization of charge collection efficiency and timewalk.
Benefits are quick computation times and that simulation inputs are derived solely from measurements as explained in the following section.
The simulation does not predict the effect of analog design modifications on the signal formation.
Future digital designs, however, can be simulated and optimized based on the realistic sensor response.

\begin{comment}
\begin{itemize}
    \item \textbf{Technology Computer-Aided Design (TCAD) with detailed geometry information.} Based on the detailed values of doping profiles, silicon resistivity as well as implant sizes the electric field in a pixel is calculated. 
    This approach is common in the design stage before production of a prototype in order to optimize the design by predicting characteristics such as capacitance, gain, timing and detection efficiency.
    Even for the experienced user it is computationally expensive, though much quicker and cheaper than fabrication and testing of a prototype. 
    \item \textbf{TCAD based on generic geometry information.} If the detailed geometry information is not disclosed by the foundry a TCAD simulation can still be possible \cite{SimGenericTCAD}.
    It relies on assumptions and careful systematic parameter variations.
    The qualitative effect of single parameter changes offers some predictive power.
    Though measurement results are beneficial to validate the assumptions entering the simulation.
    \item \textbf{A parametric fast simulation based on measurement results.} 
    A fast simulation is proposed in this work based on a charge collection parameterization obtained from measurements of charge collection efficiency and timewalk. 
    While this is obviously only feasible after sensor prototyping it is a useful and efficient tool for parameterizing the signal formation.
    The digital sensor logic as well as larger detector designs can be investigated and predicted based on this.
\end{itemize}
\end{comment}

\setcounter{page}{1} % to have first text page with number 1

\section{Charge collection measurement}
\label{sec:results}

\begin{figure}
\centering 
\includegraphics[width=0.65\textwidth]{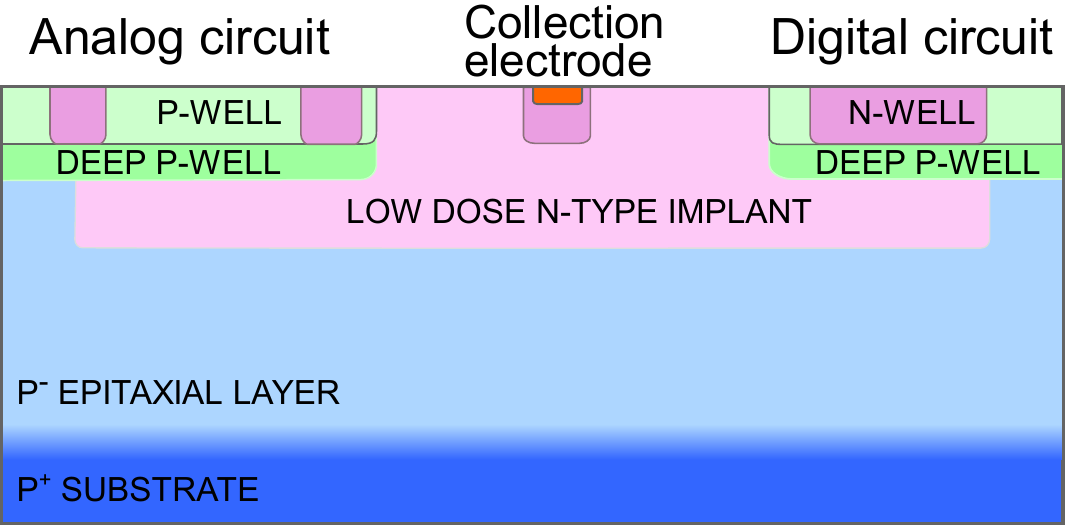}
\caption{\label{fig:Cross-section}Cross-section through the center of a MALTA2 pixel on \SI{30}{\micro\meter} epitaxial silicon.}
\end{figure}
This section describes the measurement and analysis procedure used to obtain charge collection information serving as input for the simulation.
For sensors with time-of-arrival (ToA) or time-over-threshold (ToT) readout, the charge collection data can be acquired directly.
In contrast, sensors with binary readout require data acquisition at various threshold settings.
A charge reconstruction method based on binary readout has been described in detail in ref.~\cite{MALTA2ChargeReco} for sensors fabricated on Czochralski substrate.
The same method is applied here to measurements of a non-irradiated MALTA2 sensor with \SI{30}{\micro\meter} thick epitaxial layer, produced in the Tower \SI{180}{\nano\meter} CMOS imaging technology of which a pixel cross-section is shown in figure~\ref{fig:Cross-section}.
The sensor comprises $224\times512$ pixels with a pitch of $36.4\times\SI{36.4}{\square\micro\meter}$.
A bias of \SI{-6}{\volt} is applied to the substrate and p-well.
All data can be collected within one day using a high-statistics beam facility, such as the CERN SPS.
A beam telescope is crucial for reconstructing tracks with a precision well below the pixel pitch.\\

\begin{figure}
\centering % \begin{center}/\end{center} takes some additional vertical space
\begin{subfigure}[b]{0.55\textwidth}
\includegraphics[trim=10 0 20 15,clip,width=\textwidth]{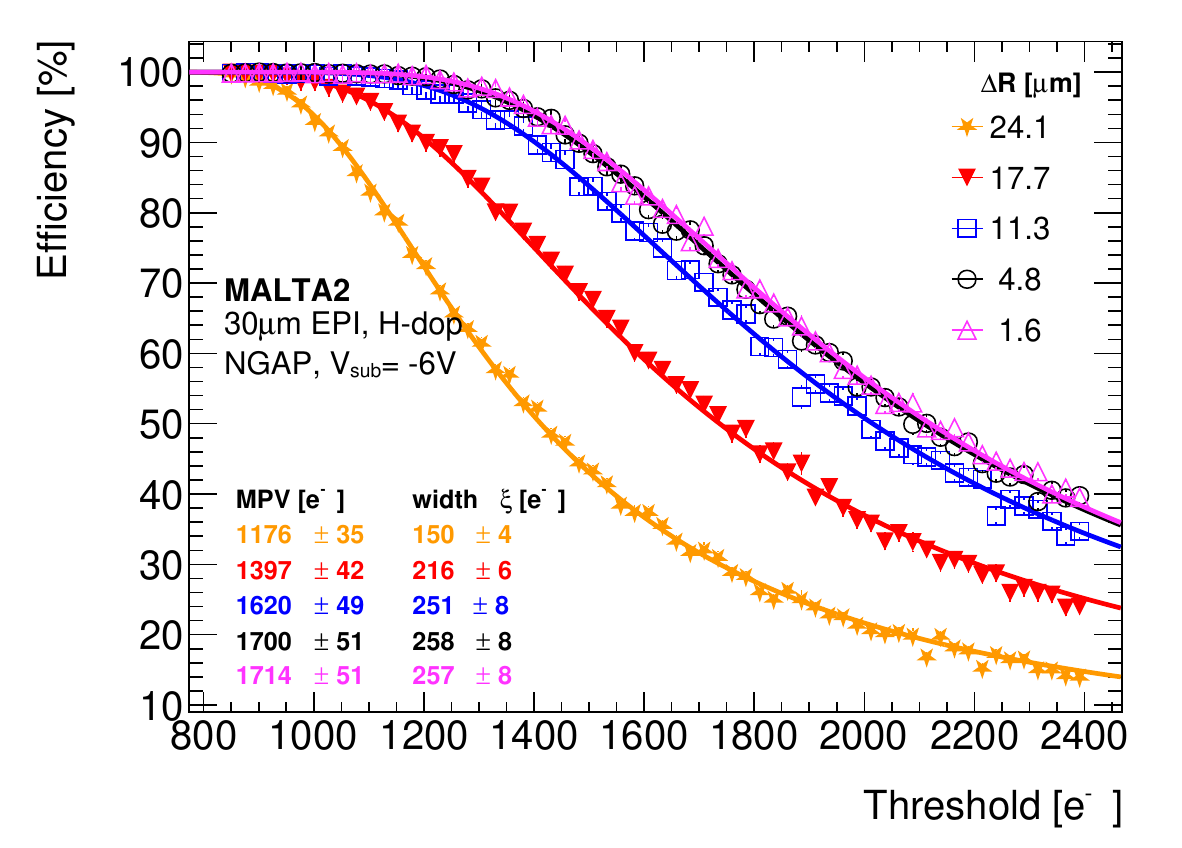}
\caption{Efficiency distribution versus threshold.}
\label{fig:Eff_Landau}
\end{subfigure}
\hfill
\begin{subfigure}[b]{0.44\textwidth}
\includegraphics[trim=0 0 0 65,clip,width=\textwidth]{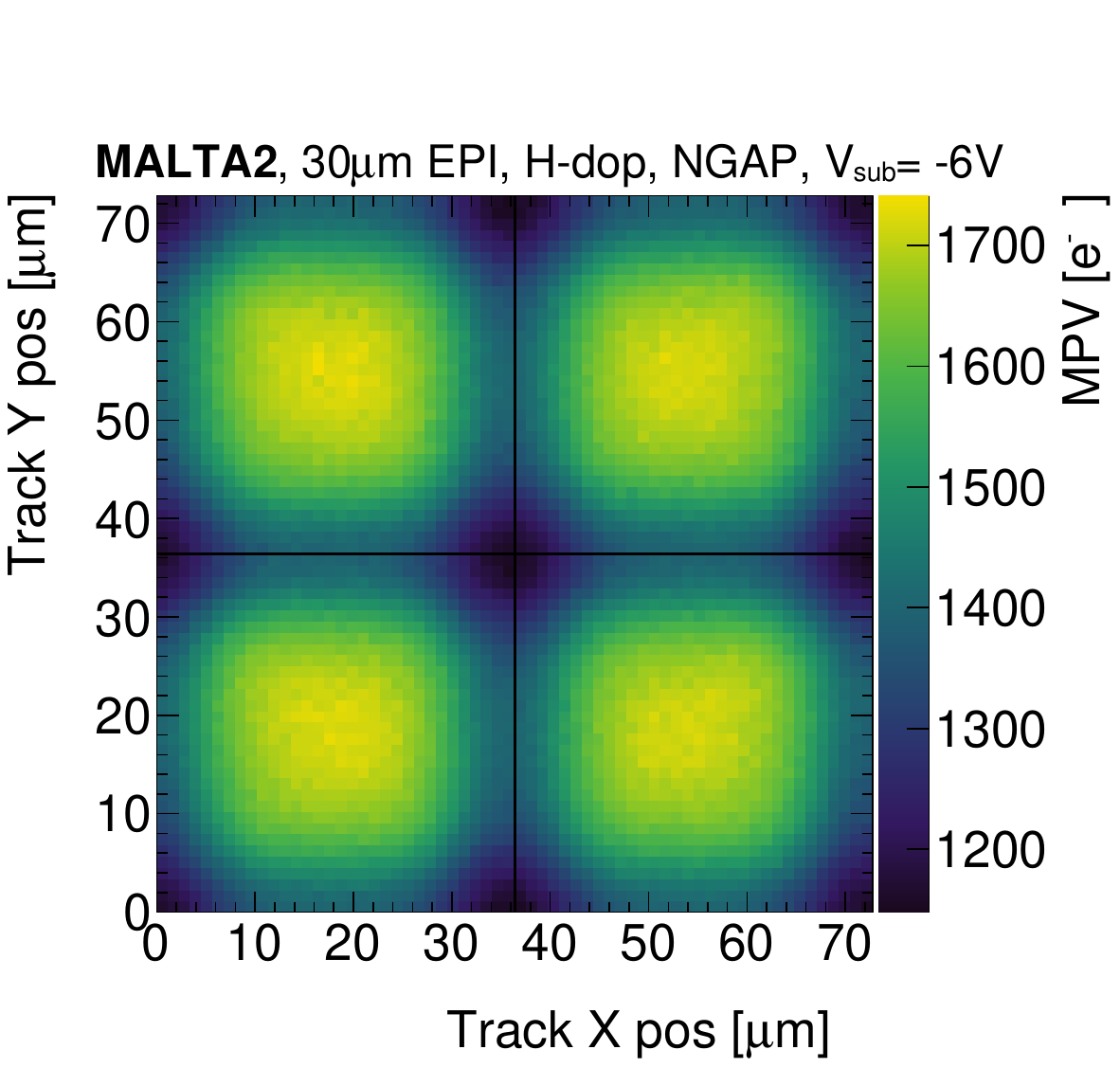} % trim top <73
\caption{\label{fig:MPV_2D}Reconstructed pixel charge.}
\end{subfigure}
\caption{\label{fig:MPV}Data of the tracking efficiency for different in-pixel regions (a) from which the most probable value (MPV) of the charge deposition is reconstructed on a $2\times2$ pixel matrix (b).}
\end{figure}

%\subsection{Most probable charge determination}
The MALTA2 sensor is characterized at the CERN SPS using a mixed hadron beam of \SI[per-mode=symbol]{180}{\giga\electronvolt\per\clight}.
It is mounted as a device under test (DUT) in the MALTA beam telescope, which consists of six tracking planes and provides a spatial resolution of $4.6 \pm \SI{0.2}{\micro\meter}$ \cite{MilouTelescope}.
The DUT is operated in a low gain setting in order to reach thresholds up to the charge deposition expected from minimum-ionizing particles.
The detection efficiency is obtained by matching telescope tracks to in-pixel regions of the DUT.
Figure~\ref{fig:Eff_Landau} shows the efficiency for different in-pixel regions of size $2.3\times\SI{2.3}{\square\micro\meter}$.
The variable $\mathrm{\Delta R}$ denotes the radial distance from the pixel center along its diagonal.
The distributions at small $\mathrm{\Delta R}$ describe the pixel center and feature larger efficiencies because charge sharing is negligible.
A fit of the upper-tail integral of the Landau distribution provides the most probable value (MPV) and width, with uncertainties including the \SI{3}{\percent} uncertainty from threshold calibration \cite{MALTA2ChargeReco,MALTA2Calibration}.
The reconstructed MPV at the pixel center, $1714\pm51\, \mathrm{e^-}$, corresponds to a sensitive silicon thickness of $29.1\pm\SI{0.8}{\micro\meter}$.
At the pixel corner (large $\mathrm{\Delta R}$), the efficiency decreases at lower thresholds, reflecting reduced collected charge due to charge sharing.
Figure~\ref{fig:MPV_2D} shows the reconstructed MPV distribution across all in-pixel regions on a $2\times2$ pixel matrix.
The charge reduction towards the pixel boundaries is due to charge sharing and a consequence of a non-uniform electric field.
The finite tracking resolution of the telescope leads to apparent charge values above \SI{50}{\percent} at the pixel boundaries and is modeled in the following section. 

\section{Charge collection model}
\begin{figure}
    \centering
    % Left image
    \begin{subfigure}[b]{0.61\textwidth}
    \vspace{0pt} % ensure true top alignment
    \includegraphics[trim=0 0 0 10,clip,width=\linewidth]{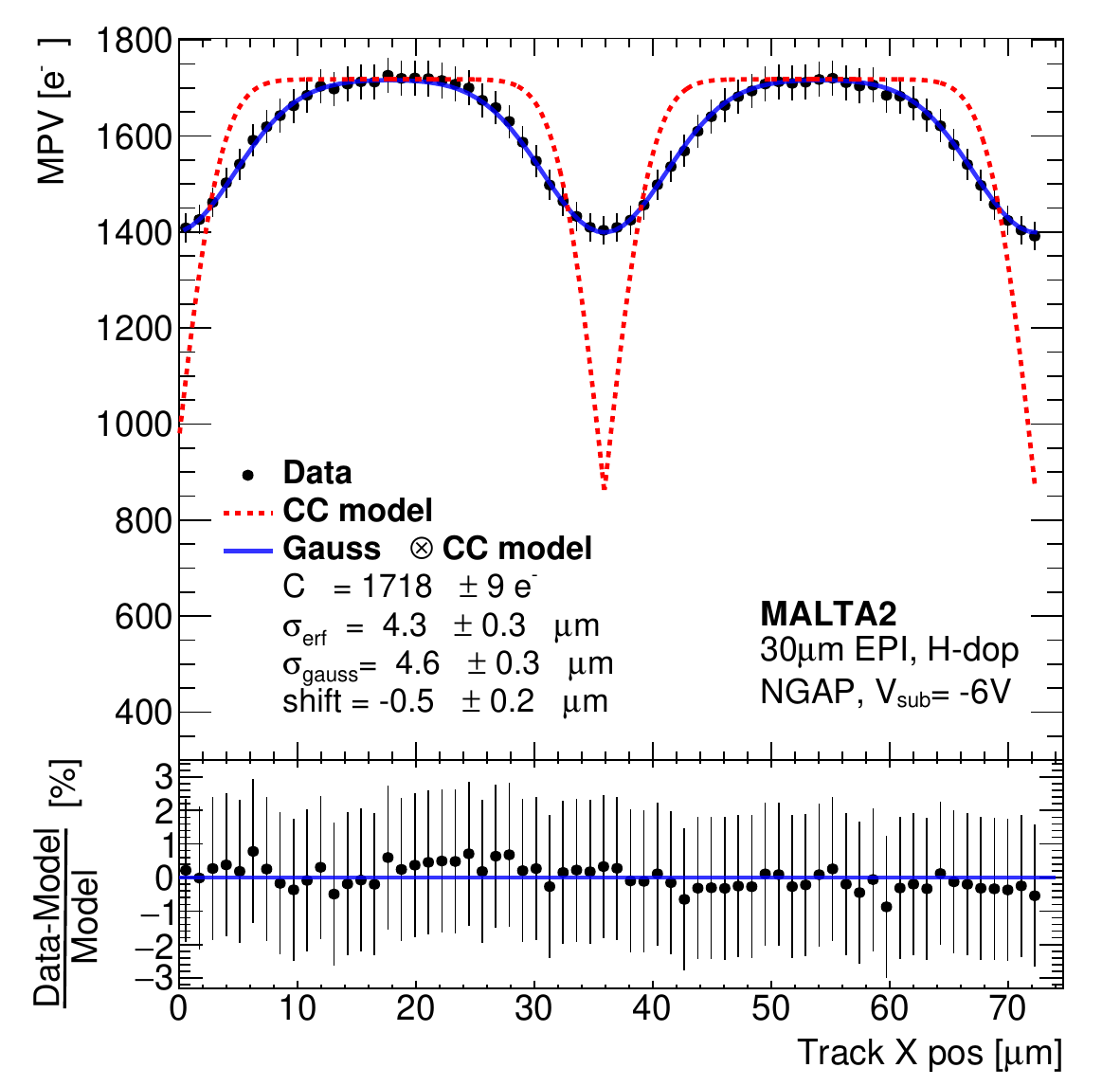}
    \caption{\label{fig:MPV_1DFit}$x$-projection of the MPV data with fit function and residuals.}
    \end{subfigure}
    %\hfill
    % Right column
    \begin{subfigure}[b]{0.37\textwidth}
    %\vspace{0pt} % ensure true top alignment
    %\qquad
    %\hspace{0.2mm}
    \includegraphics[width=0.86\linewidth]{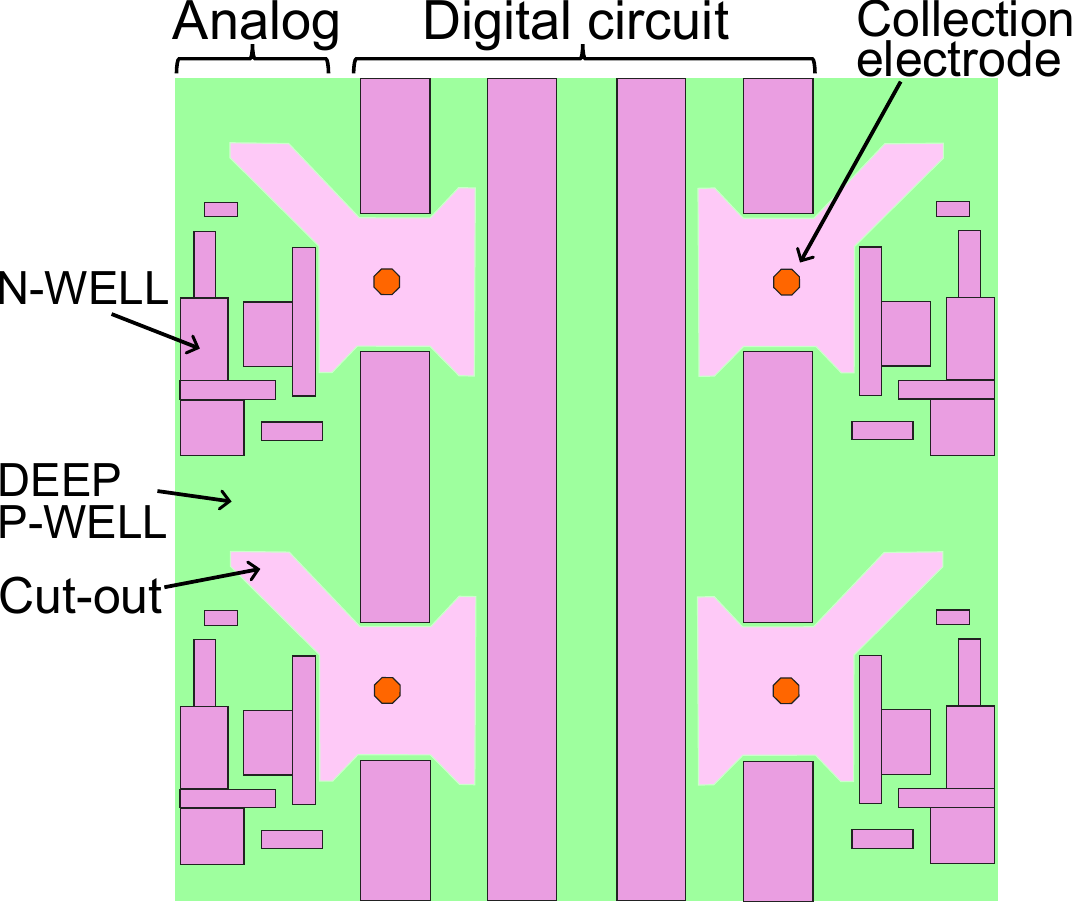}
    %\hspace{7.1mm}
    %\includegraphics[width=0.64\linewidth]{plots/oldplots/2x2Matrix_MALTA2_AnalogFE_PWELL_labeled.pdf}
    \caption{\label{fig:2DPixDesign}Simplified $2\times2$ pixel layout.}
     \vspace{0mm}
    %\qquad
    \includegraphics[trim=0 8 0 90,clip,width=\linewidth]{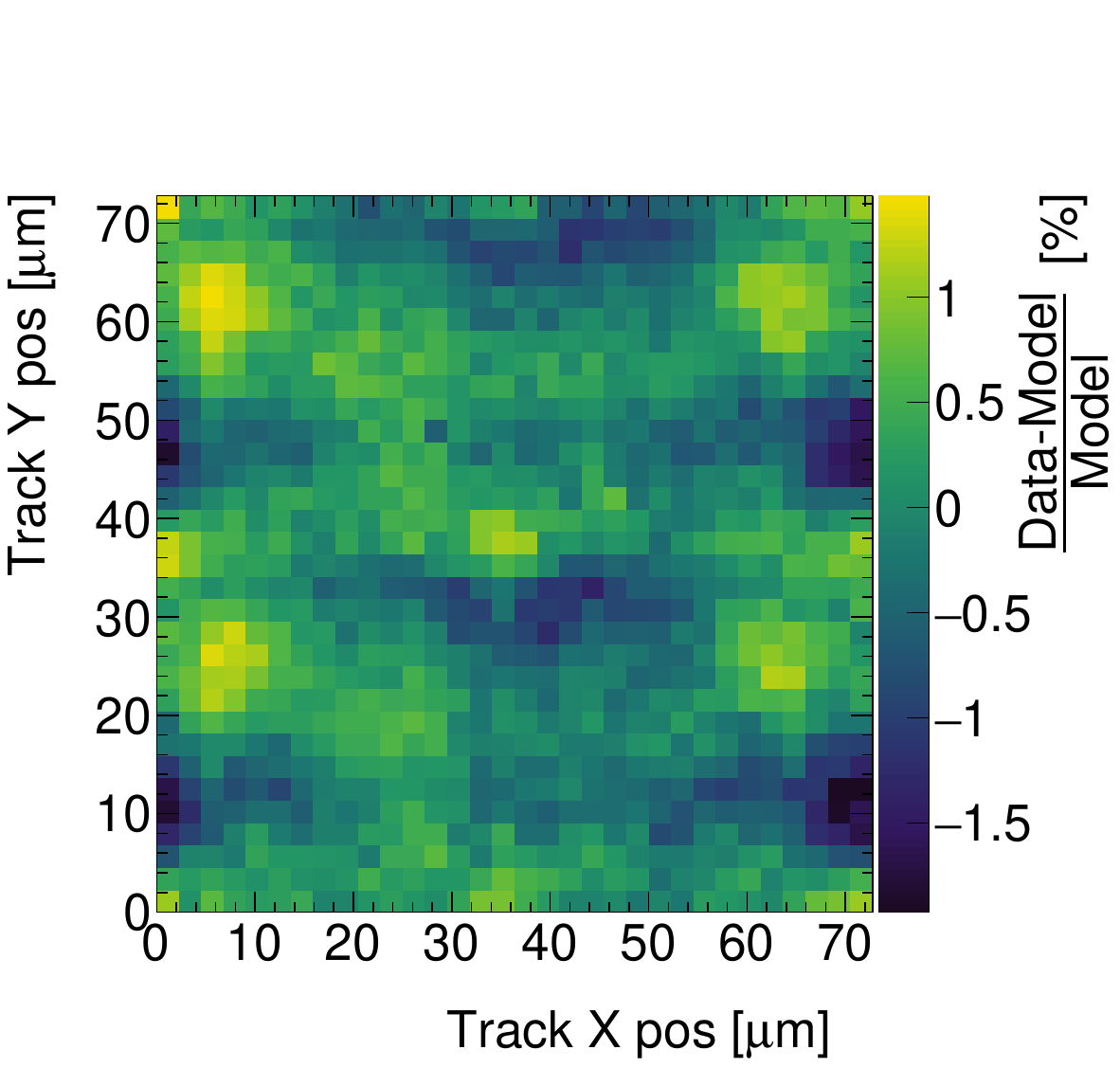}
    \caption{\label{fig:2DResid}Residuals on $2\times2$ pixel matrix.}
    \end{subfigure}
    \caption{Most probable value of the energy loss (MPV) along an X-projection in (a). 
    Shown is the one-dimensional charge collection model (red dashed) that is convoluted with a gaussian (blue line) to parameterize the data.
    The $2\times2$ pixel design in (b) explains the distribution of the two-dimensional residuals in (c).}
\end{figure}

%\subsection{Charge parameterization}
\label{sec:ChargeModel}
The reconstructed charge distribution of the MALTA2 DUT, shown in figure~\ref{fig:MPV_2D} on a $2\times2$ pixel matrix, is parameterized by an analytical charge collection model.
A one-dimensional projection through the pixel center along the $x$-axis at $y=\SI{18.2}{\micro\meter}$ is displayed in figure~\ref{fig:MPV_1DFit}.
The model, shown as a red dashed line, is periodic and symmetric, with 50\% charge sharing at the pixel boundaries.
It is defined as
\begin{equation}
\label{eq:CC}
\mathrm{CC}(x;C,\sigma_{\mathrm{erf}},\mathrm{shift}) = \frac{C}{2}\left[\mathrm{erf}\left(\frac{x-\mathrm{shift}}{\sigma_{\mathrm{erf}}}\right)  - \mathrm{erf}\left(\frac{x-\mathrm{shift}-\mathrm{pitch}}{\sigma_{\mathrm{erf}}}\right)\right]. % , x \in [\mathrm{shift},s+p).
\end{equation}
The constant factor $C$ quantifies the charge at the pixel center where charge sharing is negligible.
The periodicity is fixed to the pixel pitch of $\SI{36.4}{\micro\meter}$.
The shift parameter accounts for tracking misalignment.
To model the finite track resolution, the function is convolved with a Gaussian of width $\sigma_{\mathrm{gauss}}$, which is fitted to the data.
A best-fit value of $\sigma_{\mathrm{gauss}} = 4.6\pm\SI{0.3}{\micro\meter}$ agrees with the telescope resolution reported in ref.~\cite{MilouTelescope}.
This smearing explains why data points near pixel boundaries exceed 50\% of the central charge.
The transition at the pixel edges is characterized by the width
\begin{equation}
    \sigma_{\mathrm{erf}} = 4.3\pm\SI{0.3}{\micro\meter}.
\end{equation}
Residuals between data and model remain below \SI{1}{\percent}.
Equivalent fit results are obtained for the $y$-axis projection.
A two-dimensional model is constructed by multiplying two one-dimensional models:
\begin{equation}
\label{eq:CC_2D}
\mathrm{CC_{2D}}(x,y;C,\sigma_{\mathrm{erf}},\mathrm{shift}_x,\mathrm{shift}_y) = \mathrm{CC}(x;C,\sigma_{\mathrm{erf}},\mathrm{shift}_x) \times \mathrm{CC}(y;1,\sigma_{\mathrm{erf}},\mathrm{shift}_y). % , x \in [\mathrm{shift},s+p).
\end{equation}
When fitted to the full two-dimensional MPV distribution of figure~\ref{fig:MPV_2D}, all parameters are consistent with the one-dimensional results.
Residuals (figure~\ref{fig:2DResid}) remain below \SI{2}{\percent} across the pixel matrix, which is below the uncertainty of the charge threshold calibration of \SI{3}{\percent}.
Substructure in the residual map correlates with the pixel layout shown in figure~\ref{fig:2DPixDesign}.
The deep p-well (green area) shields the front-end electronics.
It features a cut-out around the octagonal n\textsuperscript{+} collection electrode as well as towards one corner where the shielding is not needed.
This introduces slight asymmetries which can be seen in the residuals as spots of higher reconstructed charge.
Consequently, the cut-out is beneficial for charge collection.
Regions beneath the n-wells of the analog circuit show reduced reconstructed charge, visible as negative residuals.
These features explain the small deviations from the symmetric model and reflect the influence of the analog circuit layout on charge collection.
The effect of the digital n-wells is less pronounced.
In summary, on \SI{30}{\micro\meter} epitaxial silicon the asymmetries are negligible compared to the calibration uncertainty.
More pronounced effects have been observed for small electrode MAPS fabricated in \SI{65}{\nano\meter} CMOS, where the asymmetry is increased due to a thinner epitaxial layer \cite{65nmTCAD}.

\section{Parametric fast simulation}
The simulation is implemented such that any deposited charge from GEANT4 \cite{Geant4} is shared among the seed pixel and the three pixels adjacent to the nearest pixel corner.
For a given deposition at position $(x,y)$ the charge fraction assigned to the seed pixel is calculated as
$\mathrm{CC_{2D}}(x,y;C=1,\sigma_{\mathrm{erf}}=\SI{4.3}{\micro\meter})$.
Fractions for the neighboring pixels are obtained by evaluating $\mathrm{CC_{2D}}$ at $x$ and $y$ shifted by one pixel pitch.
By definition, the sum of all four fractions adds up to unity.
This implementation restricts charge sharing to clusters of up to four pixels for a single energy deposition; however, larger clusters can occur from multiple ionization events, such as those produced by delta-rays.
The model is computationally efficient since signal formation reduces to evaluating a simple analytical function.
The same method can be extended to more complex charge collection models that incorporate charge sharing to more than four pixels, asymmetries in $x$ and $y$ or a dependence on the $z$-axis.
In this work, the model is kept as simple as possible and is validated against data.

\section{Data to simulation comparison}
\begin{figure}
    \centering
    % Left image
    \begin{subfigure}[t]{0.32\textwidth}
    \includegraphics[trim=0 0 0 50,clip,width=\linewidth]{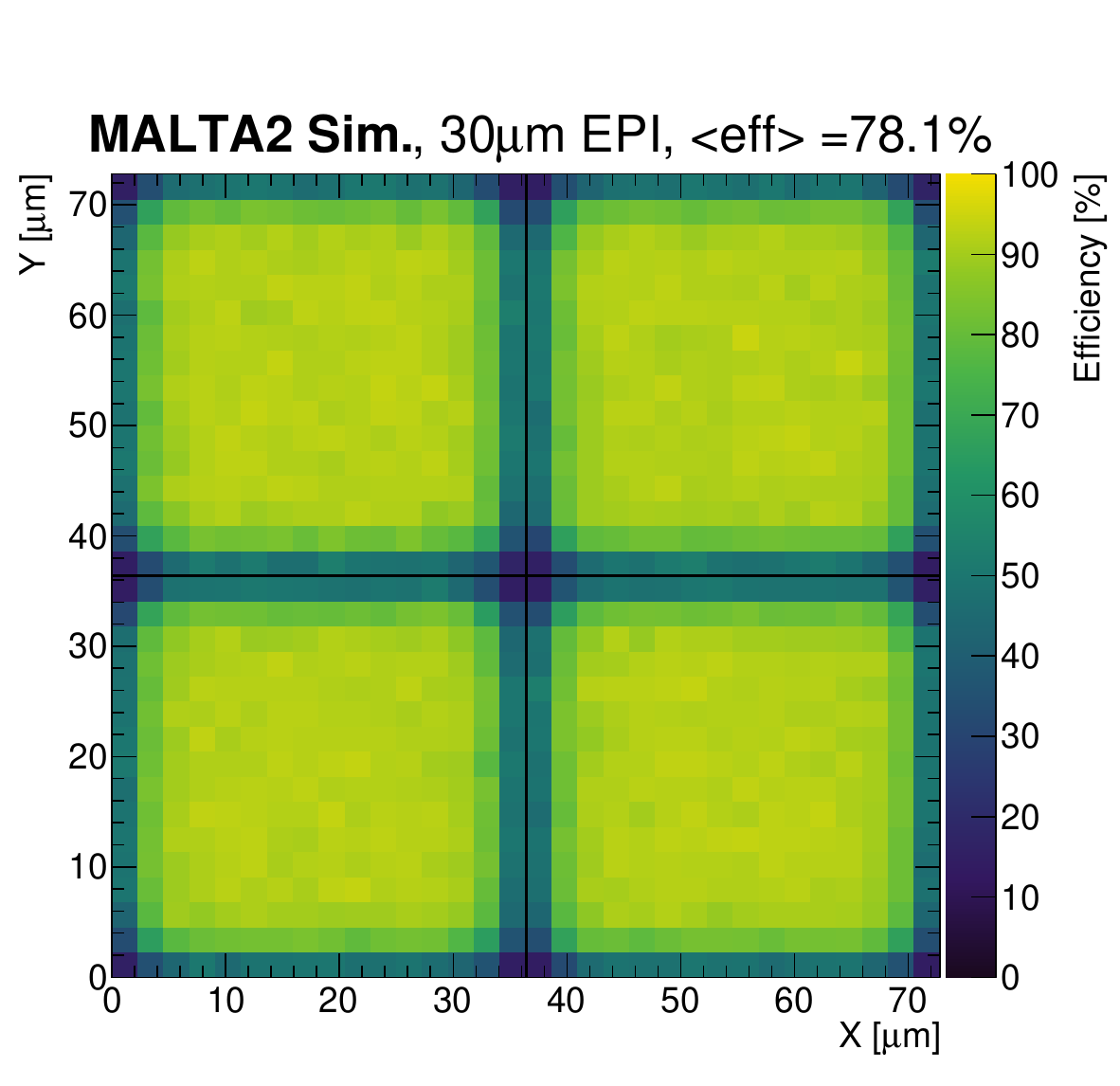}
    \caption{\label{fig:2DEff_SimUnconv}Simulation with Monte Carlo true position.}
    \end{subfigure}
    \hfill
    \begin{subfigure}[t]{0.32\textwidth}
    %\vspace{0pt} % ensure true top alignment
    %\qquad
    \includegraphics[trim=0 0 0 50,clip,width=\linewidth]{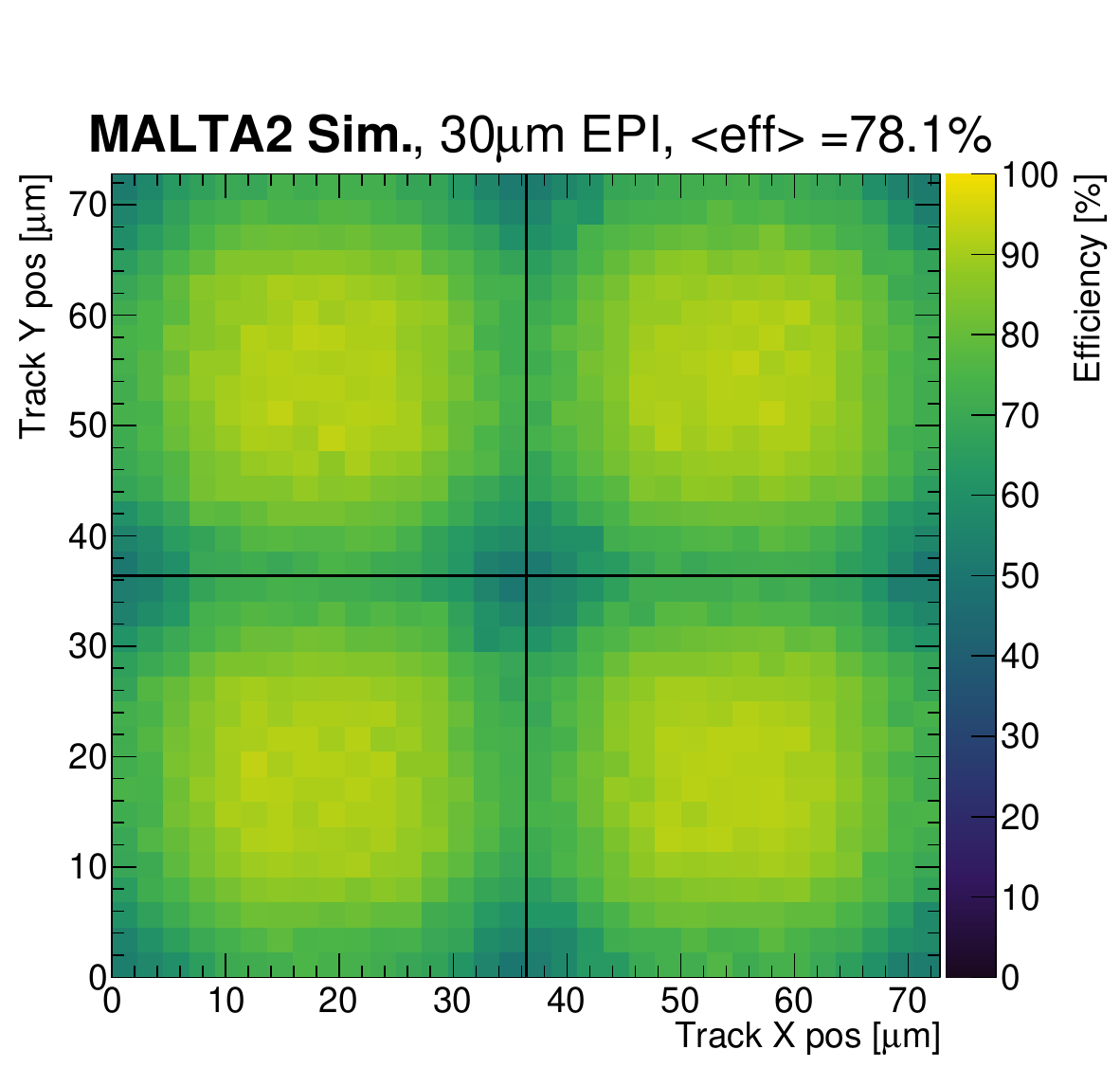}
    \caption{\label{fig:2DEff_SimTrackConv}Simulation with tracking uncertainty.}
    \end{subfigure}
    \hfill
    %\qquad
    \begin{subfigure}[t]{0.32\textwidth}
    \includegraphics[trim=0 0 0 50,clip,width=\linewidth]{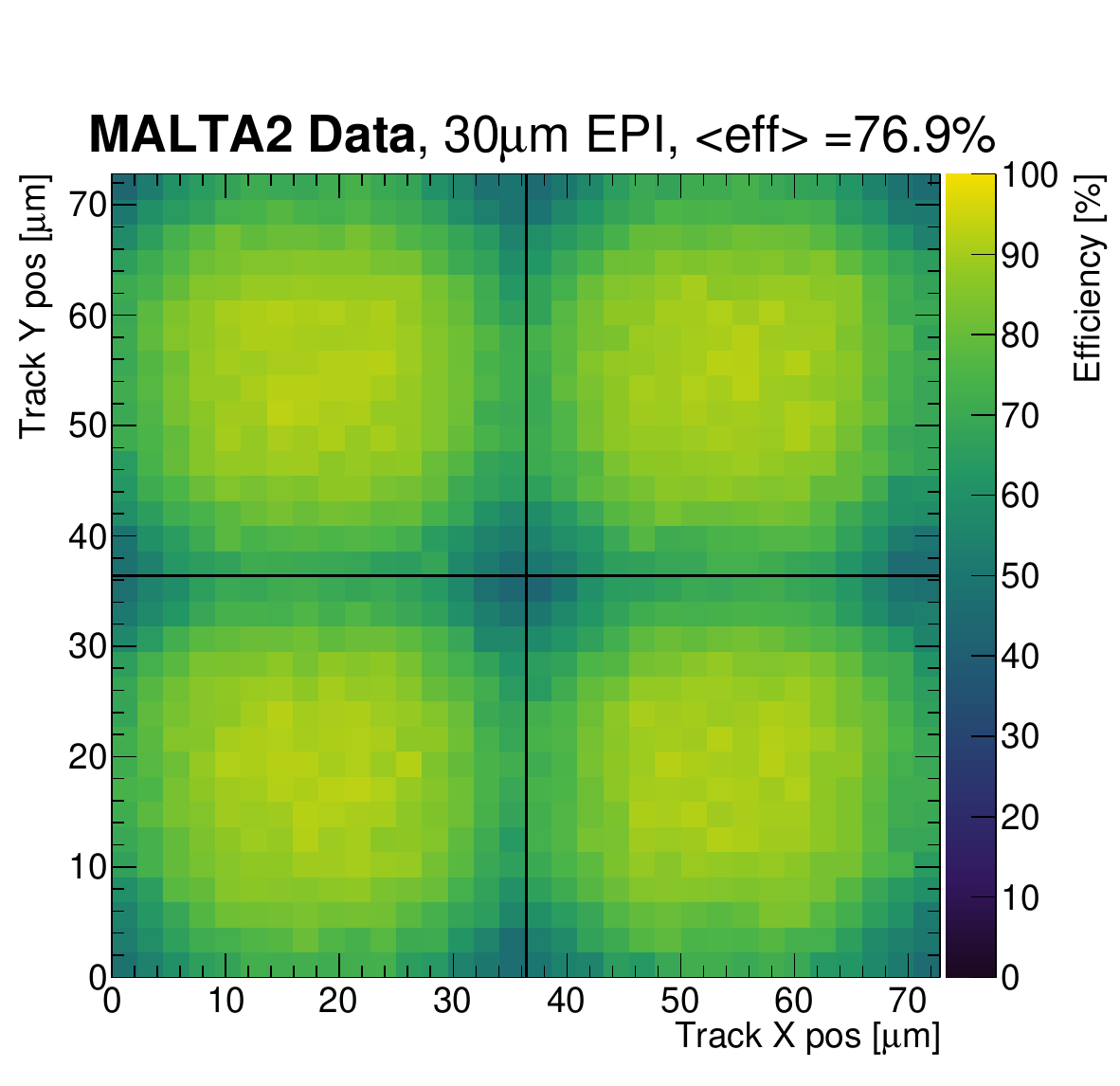}
    \caption{\label{fig:2DEff_Data}Data with tracks reconstructed with beam telescope at SPS.}
    \end{subfigure}
    \caption{\label{fig:2DEff_DataSim}In-pixel efficiency for simulation and data at a threshold of $1400\,\mathrm{e^-}$.
    The simulation matches the data when considering the tracking uncertainty of the telescope $\sigma_{\mathrm{gauss}}=\SI{4.6}{\micro\meter}$.}
\end{figure}

\begin{figure}
\centering % 
\begin{subfigure}[b]{0.49\textwidth}
\includegraphics[trim=12 0 0 15,clip,width=\textwidth]{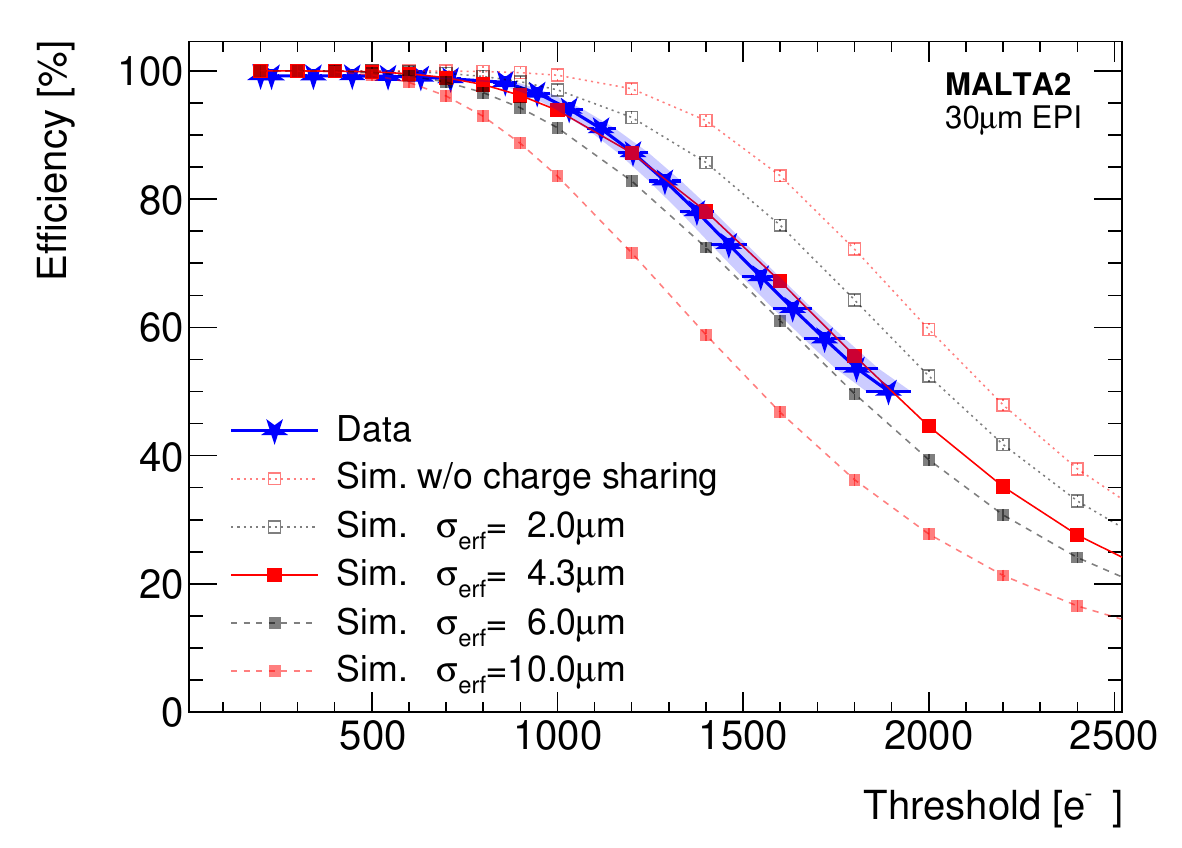}
\caption{\label{fig:AvEffDat_Sim}Average efficiency distribution versus threshold.}
\end{subfigure}
\hfill
\begin{subfigure}[b]{0.49\textwidth}
\includegraphics[trim=12 0 0 15,clip,width=\textwidth]{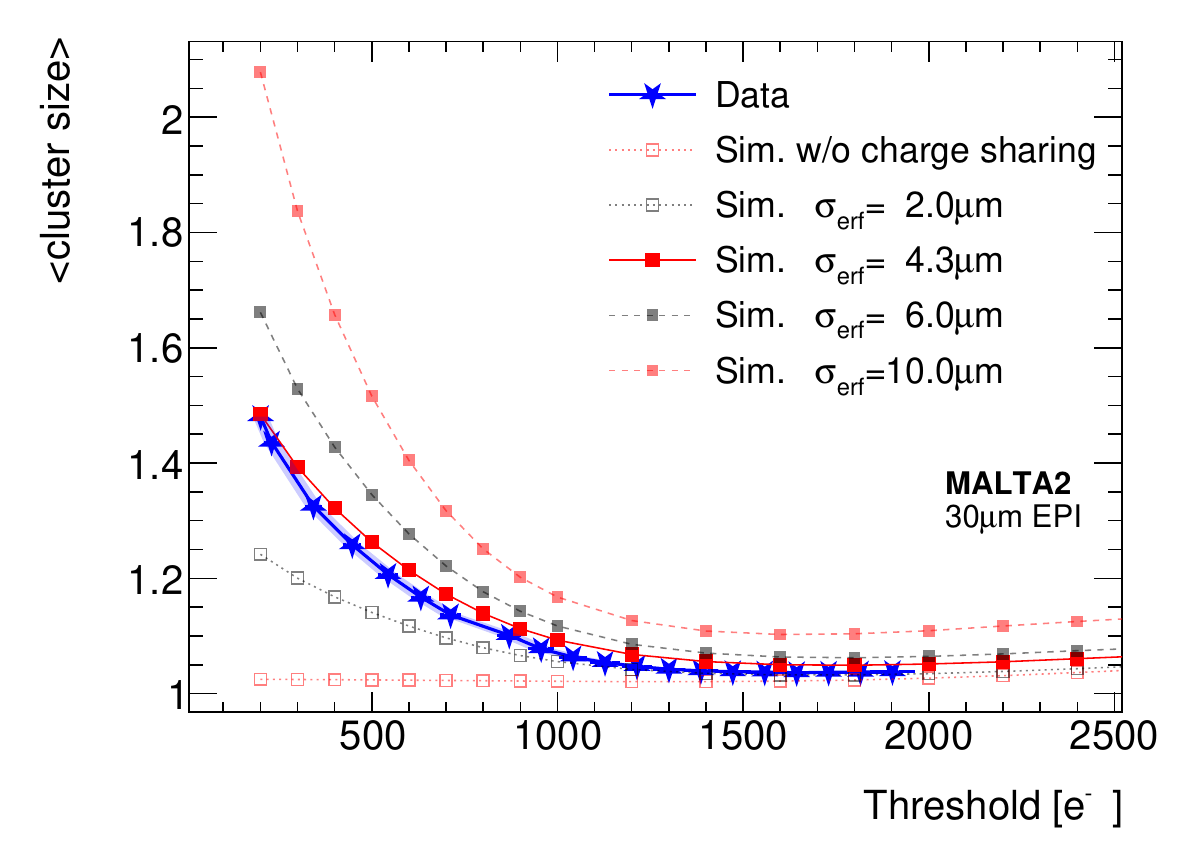} 
\caption{\label{fig:AvClSizeDat_Sim}Average cluster size distribution versus threshold.}
\end{subfigure}
\caption{\label{fig:DataSimComp}Simulation results (boxes) are compared to data (blue stars).
The model as determined in section~\ref{sec:ChargeModel} with $\sigma_{\mathrm{erf}}=\SI{4.3}{\micro\meter}$ best describes the data.}
\end{figure}

The simulated in-pixel efficiency is compared to measurements at a threshold of $1400\,\mathrm{e^-}$ in figure~\ref{fig:2DEff_DataSim}.
Figure~\ref{fig:2DEff_SimUnconv} shows the efficiency derived from the Monte Carlo true hit position, while figure~\ref{fig:2DEff_SimTrackConv} includes Gaussian smearing with a width of \SI{4.6}{\micro\meter} to emulate the telescope resolution.
This reproduces the data (figure~\ref{fig:2DEff_Data}) within an offset of approximately 1\% per bin.
The uncertainty on the mean threshold in data of $42\,\mathrm{e^-}$ is sufficient to explain this variation.
Similar or better agreement is observed for other thresholds.
Figure~\ref{fig:DataSimComp} verifies the average efficiency and cluster size across a wide threshold range.
The simulation based on eq.~\eqref{eq:CC} with $\sigma_{\mathrm{erf}} = \SI{4.3}{\micro\meter}$ accurately reproduces the measured efficiency within its \SI{3}{\percent} calibration uncertainty.
Additionally, the matching cluster size justifies the charge sharing model.
In contrast, the simulation neglecting charge sharing clearly overestimates the efficiency and underestimates the cluster size.
Also, alternative $\sigma_{\mathrm{erf}}$ values (dashed lines) fail to match the data.
This confirms the sensitivity to the model’s width parameter and validates the charge collection description established in section~\ref{sec:ChargeModel}.

\begin{comment}
    
\begin{figure}
\centering
\includegraphics[width=.7\textwidth]{plots/Comparison_Data_Sim_AvEff.pdf}
\caption{\label{fig:AvEffDat_Sim}Average efficiency distribution versus threshold for data (blue stars) and simulation (boxes).
The model as determined in section~\ref{sec:ChargeModel} with $\sigma_{\mathrm{erf}}=\SI{4.3}{\micro\meter}$ best describes the data.}
\end{figure}

\end{comment}

\section{Conclusion}
A measurement-driven parametric simulation has been developed for the MALTA2 sensor fabricated on a \SI{30}{\micro\meter} epitaxial silicon layer.
The approach provides a fast and efficient alternative to TCAD simulations, requiring no proprietary process information.
By parameterizing the charge collection behavior from beam test data, the model accurately reproduces the measured in-pixel efficiency within experimental uncertainties across the full threshold range.
The method enables rapid optimization of digital front-end design and large-scale detector architectures for demanding physics environments.
Future work involves studying digital designs for their performance in terms of hit-rates, multiple thresholds or clustering on-chip for the application in tracking and calorimetry.

\acknowledgments
This project has received funding from the European Union's Horizon 2020 Research and Innovation programme under Grant Agreement numbers: 101004761 (AIDAinnova), 675587 (STREAM), 654168 (AIDA-2020).

\end{document}